\documentclass{article}
\usepackage{amsmath,amsthm}
\usepackage{amssymb,latexsym}
\usepackage[mathscr]{eucal}
\usepackage{setspace}
\usepackage{graphics}
\usepackage{array}

\newcommand{\pa}{\partial}
\newcommand{\lp}{\left(}
\newcommand{\rp}{\right)}
\newcommand{\lb}{\left[}
\newcommand{\rb}{\right]}
\newcommand{\be}{\begin{equation}}
\newcommand{\ee}{\end{equation}}
\newcommand{\na}{\nabla}
\newcommand{\qa}{\lp {\bf q} + \frac{e {\bf A}}{m c} \rp}

\begin{document}

\title{{\bf Impulse Formulations of Hall \\Magnetohydrodynamic Equations}}         
\vspace{.5in}

\author{Bhimsen K. Shivamoggi\footnote{\large Permanent Address: University of Central Florida, Orlando, FL 32816-1364}\\J.M. Burgers Centre and Fluid Dynamics Laboratory\\Department of Physics\\Eindhoven University of Technology \\5600 MB Eindhoven, The Netherlands}        

\date{}

\maketitle

\vspace{.5in}

\noindent\Large\textbf{Abstract}

\large Impulse formulations of Hall magnetohydrodynamic (MHD) equations are developed. The Lagrange invariance of a generalized ion magnetic helicity is established for Hall MHD. The physical implications of this Lagrange invariant are discussed. The discussion is then extended to compressible Hall MHD. Compressibility effects are shown not to affect, as to be expected, the physical implications of the generalized ion magnetic helicity Lagrange invariant.

\pagebreak

\noindent\Large\textbf{1 Introduction}

\vspace{.3in}

\large Impulse formulations of magnetohydrodynamic (MHD) equations were considered by Kuzmin [1]. This led to the result, upon the use of an appropriate gauge condition, that the magnetic helicity, which is important for the study of topological properties of magnetic field lines (Moffatt [2]), is a Lagrange invariant.\\

In a high-$\beta$ plasma, on length scales in the range $d_e < l < d_i$, where $d_s \equiv c/\omega_{ps}$, $s = i, e$, is the skin depth ($i$ being reference to ions and $e$ to electrons), the electrons decouple from the ions and this results in an additional transport mechanism for the magnetic field via the Hall current (Sonnerup [3]) which is the ion-inertia contribution in generalized Ohm's law. The Hall effect leads to the generation of whistler waves whose -\\

* frequency lies between the ion-cyclotron and electron-cyclotron frequencies $\omega_{ci}$ and $\omega_{ce}$, \indent respectively;\\

* phase velocity exceeds that of Alfv\'{e}n waves for wavelengths parallel to the applied \indent magnetic field less than $d_i$.\\

\noindent Further, the decoupling of ions and electrons in a narrow region around the magnetic neutral point (where the ions become unmagnetized while the electrons remain magnetized) allows for rapid electron flows in the ion dissipation region and hence a faster magnetic reconnection process in the Hall MHD regime (Mandt et al [4]).\\

In this paper, we develop impulse formulations of Hall MHD equations and establish the Lagrange invariance of a generalized ion magnetic helicity for Hall MHD and explore its physical implications. We then extend this discussion to compressible Hall MHD and explore the robustness of the physical implications of the generalized ion magnetic helicity Lagrange invariant.

\vspace{.3in}

\noindent\Large\textbf{2 The Hall MHD Equations}

\vspace{.3in}

\large The Hall MHD equations (which were actually formulated by Lighthill [5] long ago following his far-sighted recognition of the importance of the Hall term in the generalized Ohm's law) are (in usual notation) -
\be\tag{1}
{\bf \rho} \lb \frac{\pa {\bf v}}{\pa t} + ({\bf v} \cdot \na) {\bf v} \rb = -\na p + \frac{1}{c}~{\bf J} \times {\bf B}
\ee
\be\tag{2}
\na \cdot {\bf v} = 0
\ee
\be\tag{3}
{\bf E} + \frac{1}{c} {\bf v} \times {\bf B} = \frac{1}{nec}~{\bf J} \times {\bf B}
\ee
along with Maxwell's equations
\be\tag{4}
\na \cdot {\bf B} = 0
\ee
\be\tag{5}
\na \times {\bf B} = \frac{1}{c}~{\bf J}
\ee
\be\tag{6}
\na \times {\bf E} = -\frac{1}{c}~\frac{\pa {\bf B}}{\pa t}
\ee
n is the number density of ions and electrons, {\bf v} is the ion fluid velocity, $\rho \equiv nm$, m is the ion mass, $p \equiv p_e + p_i$, $p_e$ and $p_i$ are the electron-fluid and ion-fluid pressures, respectively.

\vspace{.3in}

\noindent\Large\textbf{3 Impulse Formulation of Hall MHD}

\vspace{.3in}

\large Put,
\be\tag{7}
{\bf q} = {\bf v} + {\bf \na} \phi.
\ee
$\phi$ is chosen so that the impulse velocity {\bf q} has a compact support if the ion vorticity has a compact support (Kuzmin [1]). Then we obtain, from equation (1),
\be\tag{8}
\frac{\pa {\bf q}}{\pa t} - {\bf v} \times ({\bf \na} \times {\bf q}) = -{\bf \na} \lp \frac{p}{\rho} + \frac{1}{2} {\bf v}^2 - \frac{\pa \phi}{\pa t} \rp + \frac{1}{\rho c}{\bf J} \times {\bf B}.
\ee

On imposing the gauge condition on $\phi$ -
\be\tag{9}
\frac{\pa \phi}{\pa t} + ({\bf v} \cdot {\bf \na}) \phi + \frac{1}{2} {\bf v}^2 - \frac{p}{\rho} = 0
\ee
equation (8) becomes
\be\tag{10}
\frac{D q_i}{D t} = - q_j \frac{\pa v_j}{\pa x_i} + \frac{1}{\rho c}({\bf J} \times {\bf B})_i.
\ee

We obtain from equations (1)-(6),
\be\tag{11}
\frac{D {\bf \omega}_i}{D t} = {\bf \omega}_j \frac{\pa v_i}{\pa x_j} + \frac{1}{\rho c}\lb {\bf \na} \times ({\bf J} \times {\bf B}) \rb_i
\ee
\be\tag{12}
\frac{D A_i}{D t} = v_j \frac{\pa A_j}{\pa x_i} - \frac{1}{n e}({\bf J} \times {\bf B})_i + \frac{\pa \chi}{\pa x_i}
\ee
\be\tag{13}
\frac{D B_i}{D t} = B_j \frac{\pa v_i}{\pa x_j} - \frac{1}{ne}\lb {\bf \na} \times ({\bf J} \times {\bf B}) \rb_i
\ee
$\chi$ being an arbitrary function.\\

We derive from equations (10)-(13),
\be\tag{14}
\frac{D}{D t} \lb \qa \cdot {\bf \Omega} \rb = \Omega_j \frac{\pa}{\pa x_j}~({\bf A} \cdot {\bf v} + \chi)
\ee
where,
\be\tag{15}
{\bf \Omega} \equiv {\bf \omega} + {\bf \omega}_c,~{\bf \omega} \equiv {\bf \na} \times {\bf v},~{\bf \omega}_c \equiv \frac{e {\bf B}}{m c}.
\ee

\pagebreak

On imposing the gauge condition on $\chi$ -
\be\tag{16}
{\bf A} \cdot {\bf v} + {\bf \chi} = 0
\ee
equation (14) leads to the generalized ion magnetic helicity Lagrange invariant -
\be\tag{17}
\qa \cdot {\bf \Omega} = const.
\ee

In order to clarify the physical interpretation of this Lagrange invariant, note that we have from equations (10) and (12), on using the gauge condition (16) -
\be\tag{18}
\frac{D}{D t}~\lp q_i + \frac{e A_i}{m c} \rp = -\lp q_j + \frac{e A_j}{m c} \rp~\frac{\pa v_j}{\pa x_i}.
\ee

On the other hand, if {\bf S} is a vector field associated with an oriented material surface element of the ion fluid {\bf S} evolves according to (Batchelor [6])
\be\tag{19}
\lb \frac{\pa}{\pa t} + ({\bf v} \cdot \na) \rb {\bf S} = (-\na {\bf v})^T {\bf S}
\ee
which is identical to the equation of evolution of $({\bf q} + e{\bf A}/mc)$, namely, (18). Therefore, the field lines of $({\bf q} + e{\bf A}/mc)$~evolve as oriented ion fluid material surface elements - the direction of~~~$({\bf q} + e{\bf A}/mc)$~is orthogonal to the material surface element {\bf S} and the length of $({\bf q} + e{\bf A}/mc)$~is proportional to the area of the material surface element {\bf S}.\\

Thus, the generalized ion magnetic helicity invariant (17) physically signifies the constancy of generalized ion magnetic flux (which is the magnetic flux augmented by the ion fluid vorticity flux) through the material surface element {\bf S}. One may even view this physical significance to provide a kind of inevitability to the generalized ion magnetic helicity invariant (17).\\

It is also of interest to note that the generalized ion magnetic helicity invariant (17) offers a new physical perspective on the Beltrami state for Hall MHD (Turner [7]) given by
\be\tag{20}
{\bf \Omega} = a{\bf v}
\ee
$a$ being a constant. Using (20), (17) leads to
\be\tag{21}
{\bf S} \cdot {\bf v} = const
\ee
which physically signifies the constancy of ion volume flux through the material surface element {\bf S}, as to be expected.

\vspace{.3in}

\noindent\Large\textbf{4 Extension to Compressible Hall MHD}

\vspace{.3in}

\large Compressible Hall MHD poses certain difficulties because the plasma pressure field is now determined by thermodynamics and hence plays a dynamical role. Some of these difficulties are resolved by assuming the plasma to be barotropic, i.e., the pasma pressure is a single-valued function of the plasma density -
\be\tag{22}
P (\rho) \equiv \int\frac{dp}{\rho}.
\ee

This assumption avoids the necessity to close the compressible Hall MHD equations by adding an equation of state and an equation of the evolution of internal energy.\\

Putting again
\be\tag{7}
{\bf q} = {\bf v} + {\bf \na} \phi
\ee
and using the barotropy condition (22), we obtain
\be\tag{23}
\frac{\pa {\bf q}}{\pa t} - {\bf v} \times ({\bf \na} \times {\bf q}) = - \na \lp P + \frac{1}{2} {\bf v}^2 - \frac{\pa \phi}{\pa t} \rp + \frac{1}{m c}~{\bf J} \times \lp \frac{{\bf B}}{n} \rp.
\ee

On imposing the gauge condition on $\phi$ -
\be\tag{24}
\frac{\pa \phi}{\pa t} + ({\bf v} \cdot \na) \phi + \frac{1}{2} {\bf v}^2 - P = 0
\ee
we obtain
\be\tag{25}
\frac{D q_i}{D t} = - q_j \frac{\pa v_j}{\pa x_i} + \frac{1}{m c}~\lp {\bf J} \times \frac{{\bf B}}{n} \rp_i.
\ee

Replacing equation (2) now by the ion mass conservation equation -
\be\tag{26}
\frac{\pa n}{\pa t} + {\bf \na} \cdot (n {\bf v}) = 0
\ee
equations (1), (3)-(6) give
\be\tag{27}
\frac{D}{D t} \lp \frac{\omega_i}{n} \rp = \frac{\omega_j}{n} \frac{\pa v_i}{\pa x_j} + \frac{1}{c} \lb {\bf \na} \times \lp {\bf J} \times \frac{{\bf B}}{n} \rp \rb_i
\ee
\be\tag{28}
\frac{D A_i}{D t} = v_j \frac{\pa A_j}{\pa x_i} - \frac{1}{e} \lp {\bf J} \times \frac{{\bf B}}{n} \rp_i + \frac{\pa \chi}{\pa x_i}
\ee
\be\tag{29}
\frac{D}{D t} \lp \frac{B_i}{n} \rp = \frac{B_j}{n} \frac{\pa v_i}{\pa x_j} - \frac{1}{e} \lb {\bf \na} \times \lp {\bf J} \times \frac{{\bf B}}{n} \rp \rb_i.
\ee

We derive from equations (25), (27)-(29)
\be\tag{30}
\frac{D}{D t} \lb \qa \cdot \frac{{\bf \Omega}}{n} \rb = \frac{\Omega_j}{n} \frac{\pa}{\pa x_j} ({\bf A} \cdot {\bf v} + \chi).
\ee

On imposing the gauge condition on $\chi$ -
\be\tag{16}
{\bf A} \cdot {\bf v} + \chi = 0
\ee
equation (30) leads to the generalized ion magnetic helicity Lagrange invariant -
\be\tag{31}
\qa \cdot \frac{{\bf \Omega}}{n} = const.
\ee

In order to clarify the physical interpretation of this Lagrange invariant, note that we have from equations (25) and (28), on using the gauge condition (16),
\be\tag{32}
\frac{D}{D t} \lp q_i + \frac{e A_i}{m c} \rp = -\lp q_j + \frac{e A_j}{m c} \rp \frac{\pa v_j}{\pa x_i}.
\ee

On the other hand, material surface element {\bf S} of the ion fluid evolves according to (Batchelor [6]) -
\be\tag{33}
\lb \frac{\pa}{\pa t} + ({\bf v} \cdot \na) \rb (n {\bf S}) = -(\na {\bf v})^T (n {\bf S})
\ee
which is identical to the equation of evolution of $({\bf q} + e{\bf A}/mc)$, namely, (32).\\

Thus, the generalized ion magnetic helicity invariant (31) physically signifies the constancy of generalized ion magnetic flux (which is the magnetic flux plus the ion fluid vorticity flux) through the material surface element {\bf S}, even for compressible Hall MHD, as to be expected.\\

It is again of interest to note that the generalized ion magnetic helicity invariant (31) also offers a new perspective on the Beltrami state for compressible Hall MHD (Mahajan et al. [8]) given by
\be\tag{34}
{\bf \Omega} = b n {\bf v}
\ee
$b$ being a constant. Using (34), (31) leads to
\be\tag{35}
(n{\bf S}) \cdot {\bf v} = const
\ee
which physically signifies the constancy of ion mass flux through the material surface element {\bf S} - this is of course inevitable since the Beltrami state (34) is steady.

\vspace{.3in}

\noindent\Large\textbf{5 Discussion}

\vspace{.3in}

\large In this paper, impulse formulations of Hall MHD equations are developed. The generalized ion magnetic helicity Lagrange invariant and its physical implications are given. The discussion is then extended to compressible Hall MHD, and compressibility effects are shown not to affect, as to be expected, the physical implications of the generalized ion magnetic helicity Lagrange invariant.

\vspace{.3in}

\noindent\Large\textbf{Acknowledgments}

\vspace{.3in}

\noindent\large This work was carried out when the author held a visiting research appointment at the Eindhoven University of Technology under the auspices of the J. M. Burgers Centre. The author is very thankful to Professor Gert Jan van Heijst for his enormous hospitality. The author is thankful to Professors Swadesh Mahajan and Leon Kamp for helpful discussions.

\pagebreak

\end{document}